\documentclass[
    ,final            
  ]
  {aipproc}

\layoutstyle{6x9}

\begin{document}

\title{Precise determination of the sigma pole location from a dispersive analysis
\footnote{To appear in the proceedings of the Workshop
    on Scalar Mesons and Related Topics, Lisbon, Portugal, 11-16 Feb 2008.}}


\author{R. Garc\'\i{}a-Mart\'\i{}n}{
  address={Dpto. F\'\i{}sica Te\'orica II, Facultad de CC. F\'\i{}sicas,
Universidad Complutense, Madrid, Spain.}
}

\author{R. Kami\'nski}{
  address={Institute of Nuclear Physics, Polish Academy of Science, Krak\'ow, Poland}
}

\author{J. R. Pel\'aez}{
  address={Dpto. F\'\i{}sica Te\'orica II, Facultad de CC. F\'\i{}sicas,
Universidad Complutense, Madrid, Spain.}
}

\begin{abstract}
We review how the use of recent precise data on kaon decays
together with forward dispersion relations (FDR) and Roy's equations
allow us to determine the sigma resonance
pole position very precisely, by using only experimental input.
In addition, we present preliminary results for 
a modified set of Roy-like equations with only one
subtraction, that
show a remarkable improvement in the precision around the $\sigma$ region.
For practical applications, these results are shown to be very well
approximated by a very simple conformal expansion.
\end{abstract}

\maketitle

\section{Introduction}

The values quoted in the Particle Data Table for the sigma mass and width,
based on both pole position and Breit-Wigner parameter determinations
are very widely spread, with an estimated mass and half width of~\cite{PDT2006}:
\begin{equation}
  \sqrt{s_\sigma}\equiv M_\sigma -i\, \Gamma_\sigma/2
\simeq (400 - 1200) -i (250 - 500)\mbox{ (MeV)}.
\end{equation}
This large uncertainty is mainly due to the fact that old 
data sets for pion-pion scattering are poor and often contradictory.
Moreover, the choice of data sets varies among different works.
To make things worse, there is quite a variety of different ways to
extrapolate the data on the real axis to the complex plane, and the
pole position of the sigma is greatly affected by model dependences.

This said, model independent techniques for extrapolating
amplitudes from the real axis onto the complex plane exist in the form of
dispersion relations, which allow us to analytically continue an
amplitude away from the real axis provided we know its imaginary
part for physical values of the energy. These dispersive techniques
have already been
successfully used for predicting the position of the sigma pole,
with a remarkable agreement among the different works:
\begin{eqnarray}
  & 440 - i\,245~\mbox{MeV}&\mbox{Dobado, Pelaez (1997)~\cite{Dobado:1996ps}}\\
  & 470\pm50 - i\,260\pm25~\mbox{MeV}&\mbox{Zhou {\em et al.} (2005)~\cite{Zhou:2004ms}}
\end{eqnarray}

In particular, there exists a dispersive representation that incorporates crossing
exactly,
written by Roy~\cite{Roy:1971tc}, which involves only the partial wave amplitudes.
Roy's equations have already been used to predict the position of the sigma pole
from the theoretical predictions of ChPT~\cite{Caprini:2005zr}, obtaining:
\begin{equation}
  \sqrt{s_\sigma}= 441_{-8}^{+16}-i\,272_{-19.5}^{+9}
\end{equation}

In addition, the data coming from the E865 collaboration at Brookhaven~\cite{Pislak:2001bf},
and especially the recently published data from NA48/2~\cite{Batley:2007zz} provide
us with very precise data on pion-pion scattering at very low energies,
These allow us to obtain very reliable parametrizations of the S0 wave at low
energy~\cite{PelaezTalk}, from which the scattering lengths can be
directly extracted~\cite{Yndurain:2007qm} with a remarkable precision and
in good agreement with the theoretical predictions of
ChPT~\cite{Caprini:2005zr}.

Our aim is thus to perform a dispersive analysis, including all available
experimental data, in order to give a precise and model independent
determination of the sigma pole position, by using exclusively data,
analyticity and crossing symmetry. We use both Forward Dispersion
Relations (FDR) and Roy's equations, without assuming ChPT,
so that we can actually test its predictions.

\section{Approach and results}

The details on the parametrizations used for the data have been explained fully
in this conference (see talk by J. R. Pel\'aez~\cite{PelaezTalk}),
as well as in Ref.~\cite{Kaminski:2008qe}
-- that we will denote
by KPY08. It is enough to say here that two different
sets of parameters are considered:
\begin{itemize}
\item {\em Unconstrained
Fits to Data} (UFD), in which each partial wave is fitted independently.
This set satisfies both FDR and Roy's equations within the experimental
errors in all waves except the Roy equation for the S2 wave, 
for which the deviation is
about $1.3\,\sigma$, and 
the antisymmetric FDR above 930 MeV by a couple of standard deviations.
\item{\em Constrained Fits to Data} (CFD), obtained
 by constraining the fits to satisfy simultaneously FDR and Roy's
equations, so that all waves are correlated. The CFD set provides
a remarkably precise and reliable description of the experimental data,
and at the same time satisfy the analytic properties remarkably well.
\end{itemize}
These two sets provide a reliable parametrization for the imaginary
part of the partial waves that we need as input for Roy's equations.

An elastic resonance has an associated pole on the second Riemann sheet of the
complex plane S-matrix, which, as it is well known, corresponds by unitarity
to a zero on the first sheet. As usual then, we just need to look numerically for
zeroes of the S-matrix on the physical sheet:
\begin{equation}
  S_0^0(s)=1+2i\sigma(s) t_0^0(s),
\end{equation}
where the analytic extension
of the partial wave amplitudes away from the real axis is given by
Roy's equations:
\begin{equation}
  t_0^0(s)=a_0^0 + \frac{s-4M_\pi^2}{12M_\pi^2}(2a_0^0-5a_0^2)
          + \sum_{I=0}^2\sum_{\ell=0}^1 \int_{4M_\pi^2}^\Lambda ds'\,
            k_{0\ell}^{0I}(s,s')\,\mbox{Im } t_\ell^I(s)
          + d_0^0(s).
\end{equation}

The domain of validity of Roy's equations has been shown to cover the
region of the complex plane where the sigma lies~\cite{Caprini:2005zr}.

Taking the UFD set as the input for Roy's equations, we find an S-matrix
zero at $\sqrt{s}=(426\pm 25)-i(241 \pm 17)$~MeV. However, Roy's equations
are not completely satisfied by this data set, thus the pole position will
be much more reliable if the input satisfies the equations,
as it is the case for the CFD set. In this case we find:
\begin{equation}
  \sqrt{s_\sigma} = (456 \pm 36) - i (256 \pm 17)\mbox{ MeV},
\end{equation}
which still has big uncertainties due to the strong dependence of Roy's
equations on the scattering lengths, in particular of the $a_0^2$, which
is known with less precision. These values are, however, subject to further
improvement and should be considered preliminary. It should also be noted
that they are in perfect agreement with the theoretical prediction
by Caprini {\em et al.} of
$\sqrt{s_\sigma}=441_{-8}^{+16}-i\,272_{-19.5}^{+9}$.

\section{Work in progress}
The three authors of this work together with F.~J.~Yndur\'ain (see J. R. Pel\'aez
talk in this conference~\cite{PelaezTalk}) have derived
a modified set of Roy-like equations which are based on once-subtracted
dispersion relations -- Roy's equations are twice subtracted.
These new equations (GKPY for brevity)
have a very different dependence on the observables.
In particular,
given the same input,
the uncertainty dependence on scattering lengths is much weaker.
Indeed, at low energy (below $\sim 350$~MeV)
we find the uncertainties for GKPY equations bigger than the ones given by Roy's
equations, but above $\sim 400$~MeV they are already smaller, as they
do not increase with energy. This allows us to obtain the position of
the sigma pole from Constrained Fits to Data
with higher accuracy than using standard Roy's equations
alone.

Moreover, we have already performed a preliminary 
 Constrained Fit to Data (CFD-II) in which these new
equations are also imposed as new constraints within errors (see~\cite{PelaezTalk}).
This 
gives rise to a new set of parameters which better encode the experimental
information together with unitarity, analyticity and crossing symmetry,
therefore allowing us to obtain a more precise and reliable determination
of the sigma pole. The result for the sigma pole
position using the very preliminary CFD-II set is:
\begin{eqnarray}
  \label{eq:sigmapole}
&&  \sqrt{s_\sigma} = (462 \pm 51) - i (264 \pm 20)\mbox{ MeV}, \;  \hbox{\rm (preliminary from Roy Eqs.)}
\\
&&  \sqrt{s_\sigma} = (458 \pm 15) - i (262 \pm 15)\mbox{ MeV}, \;  \hbox{\rm (preliminary from GKPY)}
\label{sigmagkpy}
\end{eqnarray}
although, as it can be seen in the Figure~\ref{fig}a, more work is needed
on the $f_0(980)$ region, as GKPY equations can discern among solutions
that were equivalent for Roy's equations. The analysis should be complete
 within the next few months.

\section{The conformal expansion}
The most rigorous way to extrapolate
to the complex plane is by using Roy's 
or GKPY equations. However, dealing with the whole set of equations is
complicated and computationally tedious. For simple applications, there exists
a simple approximate solution, which is very easy to handle: the
conformal expansion. This is a model independent
parametrization of the experimental data at low energies, based on unitarity
and elasticity, and can describe experimental data with few parameters.
The explicit details are explained in full length in Ref.~\cite{Yndurain:2007qm},
here it suffices to remember that, for elastic scattering, a given partial
amplitude of definite isospin $I$ and angular momentum $\ell$ can be written
as $t_\ell^I(s)=\frac{1}{\psi(s)-i\sigma(s)}$, with $\psi(s)$ being the
effective range function, which can be series expanded in the conformal
variable $\omega(s)=\frac{\sqrt{s}-\sqrt{s_0-s}}{\sqrt{s}+\sqrt{s_0-s}}$
as follows:
\begin{equation}
  \psi(s)=\frac{M_\pi^2}{s-z_0^2/2}
  \left\{ \frac{z_0^2}{M_\pi \sqrt{s}} + B_0 + B_1\omega(s) + B_2\omega(s)^2 + \dots
    \right\}
\end{equation}
Three parameters are enough to describe the experimental data below
the $\pi\pi\rightarrow\bar KK$ inelastic threshold in all studied $\pi\pi$
partial waves ($S$, $P$, $D$, $F$ and $G$). 

\begin{figure}
  \label{fig}
  \includegraphics[width=.55\textwidth,height=.25\textheight]{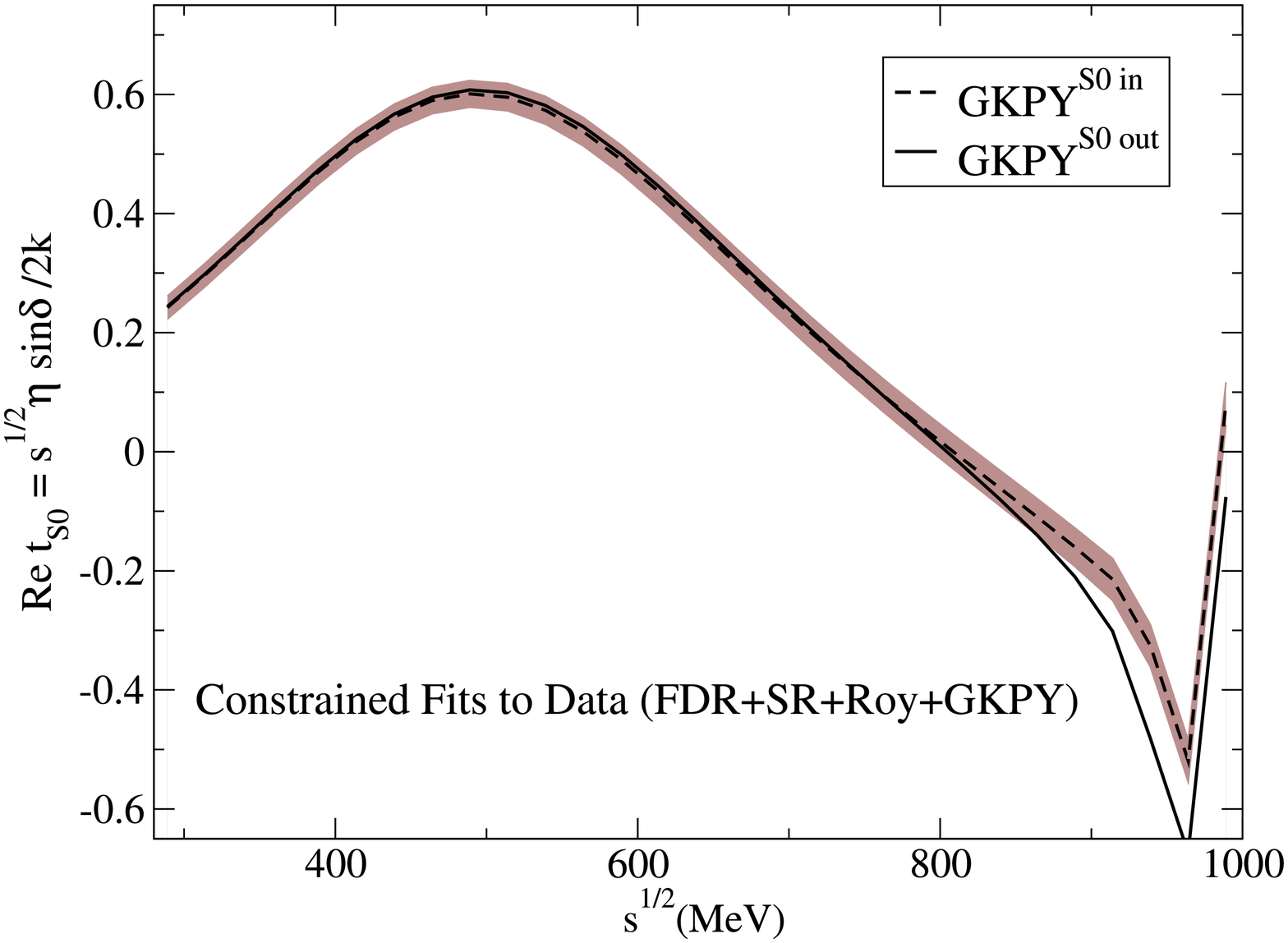}
  \includegraphics[width=.45\textwidth,height=.25\textheight]{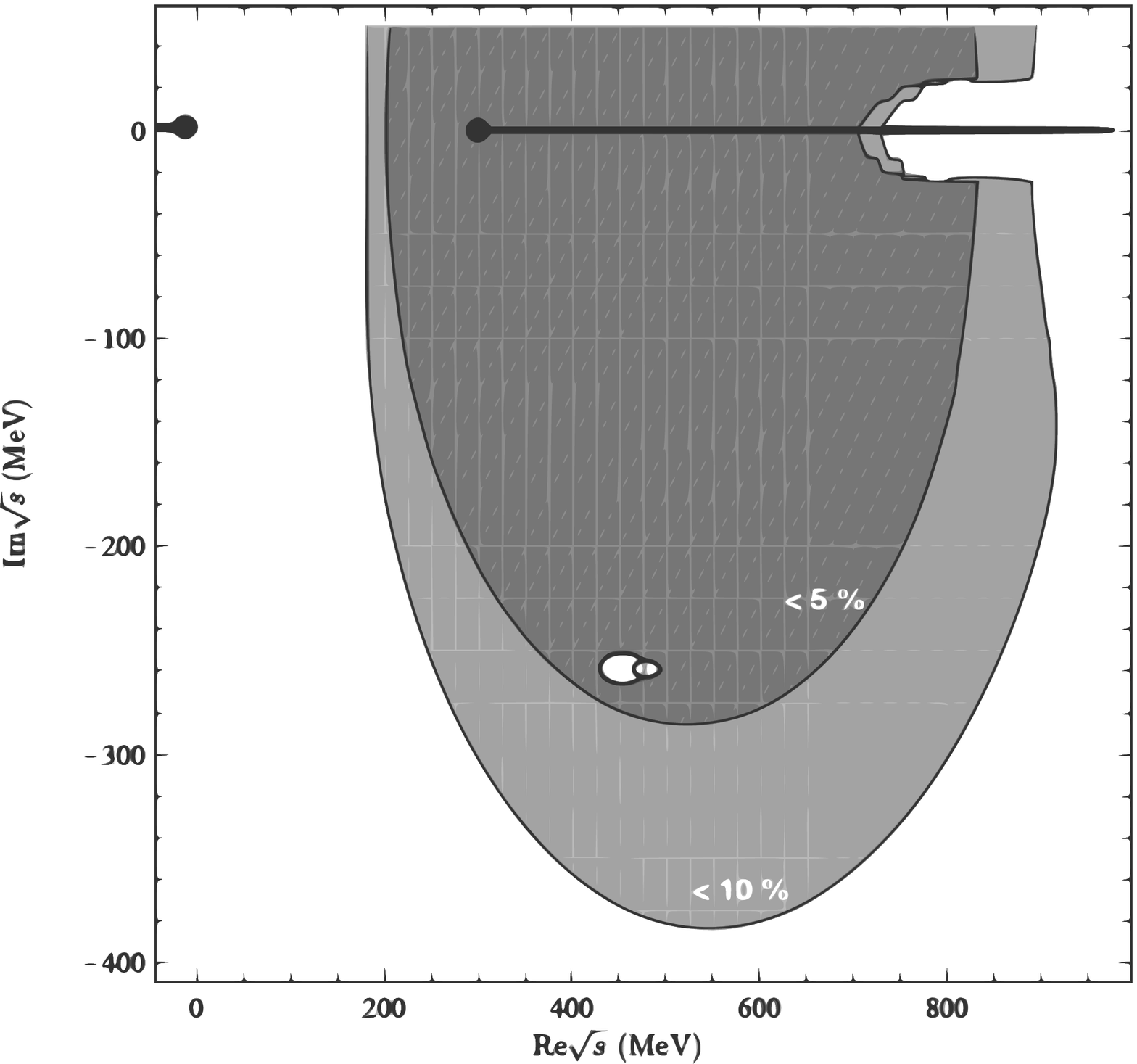}
  \caption{
a) Difference between input and output of GKPY equations for
$\mbox{Re }t_0^0$, with the corresponding error band. The smaller
uncertainties carried by these new equations will allow us to improve
the fit in the $f_0(980)$ region.
b) Relative difference between the conformal expansion and Roy's
equations. The contours show the regions inside of which the difference is
less than 5 and 10 percent. The sigma pole positions with their errors
are plotted as ellipses.}
\end{figure}

We show in
Figure~\ref{fig}b the complex plane for $\sqrt{s}$, plotting the
contours inside of which the 
relative difference between the amplitude
calculated with GKPY equations and calculated with
 the conformal expansion, i.e.,
$  \Delta_{00}(s)=
  \frac{\left\vert t_{00}^{conf}(s)-t_{00}^{GKPY}(s)  \right\vert}
  {\frac{1}{2} \left\vert t_{00}^{conf}(s)+t_{00}^{GKPY}(s)  \right\vert},
$
is less than 5 or 10 percent. Note that we use our preliminary parameters obtained
by constraining the data fit to satisfy FDR, Roy's equations and GKPY
equations. We can see that for the sigma region, {\it the difference between
both calculations is less than 5 percent}, and indeed the pole locations
overlap within their errors: using the same input as for Eq.~(\ref{eq:sigmapole})
we find:
\begin{equation}
  \sqrt{s_\sigma^{conf}} = (478 \pm 17) - i\,(262 \pm 7)~\mbox{MeV}.
\label{sigmaconf}
\end{equation}
We see that the systematic uncertainties associated with
the conformal expansion, which correspond to neglecting
crossing symmetry and the inelastic cut---$f_0(980)$ region---
are bigger than our previous
estimation in Ref.~\cite{Yndurain:2007qm}, but still
within the 5 percent band.
The crudeness of our estimation of systematic uncertainties
 has been pointed out by I. Caprini in this conference \cite{Caprini:2008bm}.
Certainly in ref.~\cite{Yndurain:2007qm} we only estimated crudely 
the systematic error from 
different parametrizations and truncations of our conformal expansion,
and we came up with $\Delta_{sys}M_\sigma=\pm11\,$MeV and 
$\Delta_{sys}\Gamma_\sigma/2=\pm 2\,$MeV. 
 I. Caprini \cite{Caprini:2008bm}, using an arbitrary sampling of 
conformal parametrizations and conformal
variables -- which are different from ours -- 
has provided a new estimate 
$\Delta_{sys}M_\sigma\simeq 40\,$MeV and 
$\Delta_{sys}\Gamma_\sigma/2=\pm 40\,$MeV, 
namely a factor of 3 or 4 larger than ours.
Nevertheless, 
the difference between central values 
for the sigma pole position obtained from GKPY equations
and our conformal expansion provides the 
systematic theoretical uncertainties in 
our conformal expansion calculation, without depending on a sampling
of conformal variables and parametrizations.
In view of the
differences between Eqs.\eqref{sigmagkpy} and \eqref{sigmaconf},
we certainly agree that the systematic uncertainty of the 
pole obtained from our conformal parametrization
 was underestimated, although not as much as it is suggested
in \cite{Caprini:2008bm} but just by a factor of two
$\Delta_{sys}M_\sigma\simeq 20\,$MeV. 

To summarize this section, up to the sigma region of the complex plane, 
our conformal expansion 
provides a very simple and reliable approximation, accurate to 5\%,
to the $\pi\pi$ scattering amplitude as obtained
from data fits constrained to satisfy FDR, Roy and GKPY equations.

\begin{theacknowledgments}
We thank Prof. F. J. Yndur\'ain for his comments and suggestions,
the organizers for creating the nice scientific atmosphere of the workshop
and the Spanish research contracts PR27/05-13955-BSCH, 
FPA2004-02602, UCM-CAM 910309 and BFM2003-00856 for partial financial support.
\end{theacknowledgments}

\bibliographystyle{aipproc}   
\bibliography{rgarciamartin}

\hyphenation{Post-Script Sprin-ger}
\begin{thebibliography}{11}
\expandafter\ifx\csname natexlab\endcsname\relax\def\natexlab#1{#1}\fi
\providecommand{\enquote}[1]{``#1''}
\expandafter\ifx\csname url\endcsname\relax
  \def\url#1{\texttt{#1}}\fi
\expandafter\ifx\csname urlprefix\endcsname\relax\def\urlprefix{URL }\fi
\providecommand{\eprint}[2][]{\url{#2}}

\bibitem[Yao et~al.(2006)]{PDT2006}
W.~M. Yao, et~al., \emph{J. Phys.} \textbf{G33}, 1--1232 (2006).

\bibitem[Dobado and Pelaez(1997)]{Dobado:1996ps}
A.~Dobado, and J.~R. Pelaez, \emph{Phys. Rev.} \textbf{D56}, 3057--3073 (1997),
  \eprint{hep-ph/9604416}.

\bibitem[Zhou et~al.(2005)]{Zhou:2004ms}
Z.~Y. Zhou, et~al., \emph{JHEP} \textbf{02}, 043 (2005),
  \eprint{hep-ph/0406271}.

\bibitem[Roy(1971)]{Roy:1971tc}
S.~M. Roy, \emph{Phys. Lett.} \textbf{B36}, 353 (1971).

\bibitem[Caprini et~al.(2006)]{Caprini:2005zr}
I.~Caprini, G.~Colangelo, and H.~Leutwyler, \emph{Phys. Rev. Lett.}
  \textbf{96}, 132001 (2006), \eprint{hep-ph/0512364}.

\bibitem[Pislak et~al.(2001)]{Pislak:2001bf}
S.~Pislak, et~al., \emph{Phys. Rev. Lett.} \textbf{87}, 221801 (2001),
  \eprint{hep-ex/0106071}.

\bibitem[Batley et~al.(2008)]{Batley:2007zz}
J.~R. Batley, et~al.  (2008), cERN-PH-EP-2007-035.

\bibitem[Pelaez(2008)]{PelaezTalk}
J.~R. Pelaez, \emph{In this conference}  (2008), \eprint{0804.2632}.

\bibitem[Yndurain et~al.(2007)]{Yndurain:2007qm}
F.~J. Yndurain, R.~Garcia-Martin, and J.~R. Pelaez, \emph{Phys. Rev.}
  \textbf{D76}, 074034 (2007).

\bibitem[Kaminski et~al.(2008)]{Kaminski:2008qe}
R.~Kaminski, J.~R. Pelaez, and F.~J. Yndurain  (2008), \eprint{arXiv:0710.1150
  [hep-ph]}.

\bibitem[Caprini(2008)]{Caprini:2008bm}
I.~Caprini, \emph{In this conference}  (2008), \eprint{0804.2108}.

\end{thebibliography}

\end{document}